\documentclass[fleqn,twoside]{article}
\usepackage{espcrc2}

\usepackage{graphicx}
\usepackage[figuresright]{rotating}
\usepackage{epsf,amssymb}

\newcommand{\be}{\begin{equation}}
\newcommand{\ee}{\end{equation}}
\newcommand{\bea}{\begin{eqnarray}}
\newcommand{\eea}{\end{eqnarray}}
\newcommand{\beas}{\begin{eqnarray*}}
\newcommand{\eeas}{\end{eqnarray*}}

\def\verttl{{\;\raisebox{-0mm}{\epsfysize=3mm\epsfbox{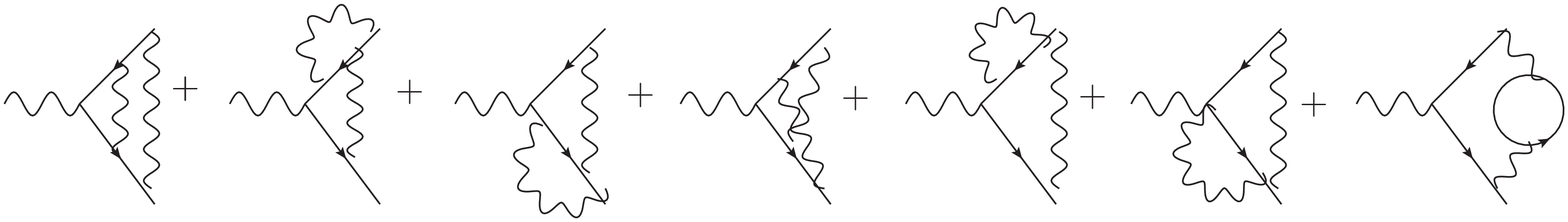}}\;}}

\def\phol{{\;\raisebox{-0mm}{\epsfysize=2mm\epsfbox{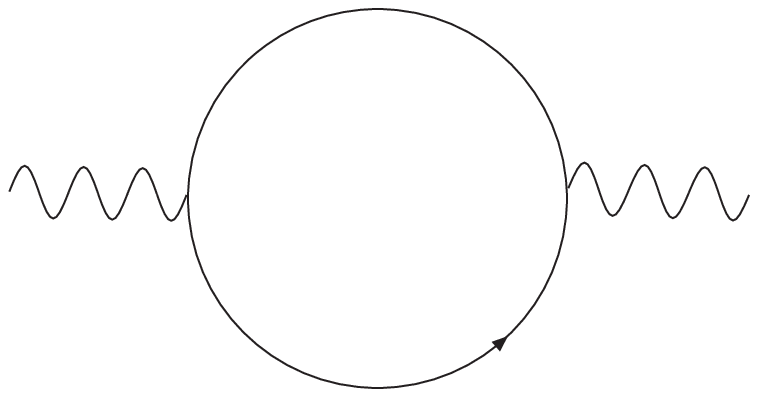}}\;}}
\def\feol{{\;\raisebox{-0mm}{\epsfysize=2mm\epsfbox{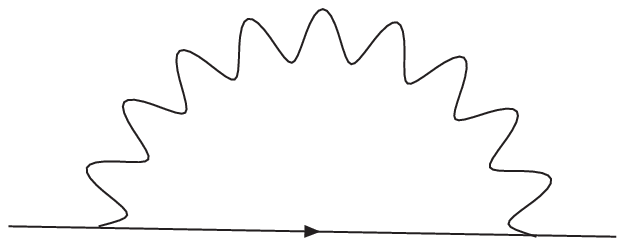}}\;}}
\def\vertol{{\;\raisebox{-0mm}{\epsfysize=2mm\epsfbox{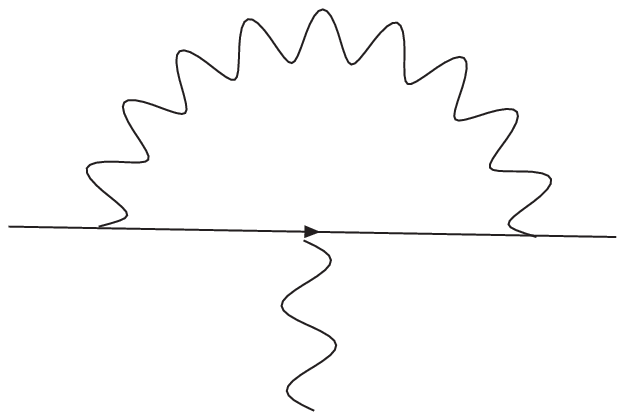}}\;}}

\def\triangle{{\;\raisebox{-3mm}{\epsfysize=6mm\epsfbox{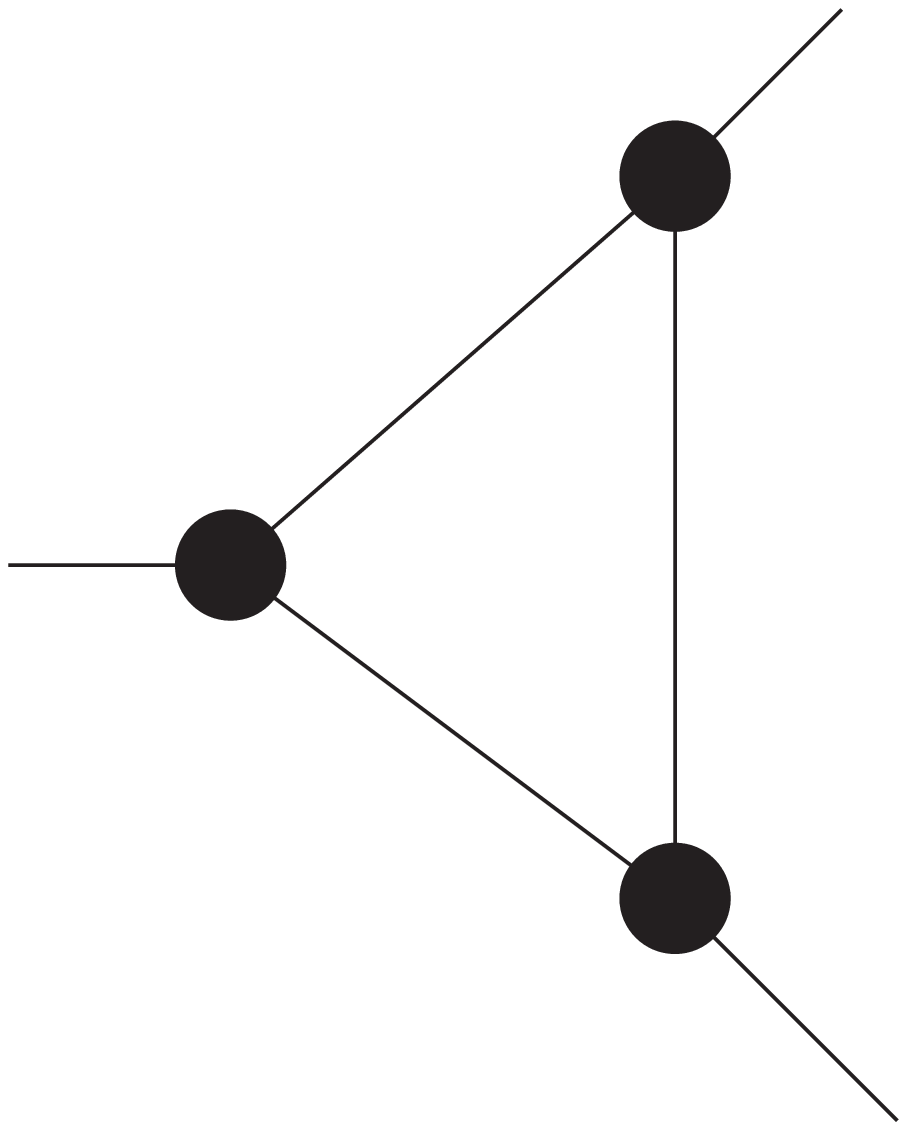}}\;}}
\def\triangleab{{\;\raisebox{-3mm}{\epsfysize=6mm\epsfbox{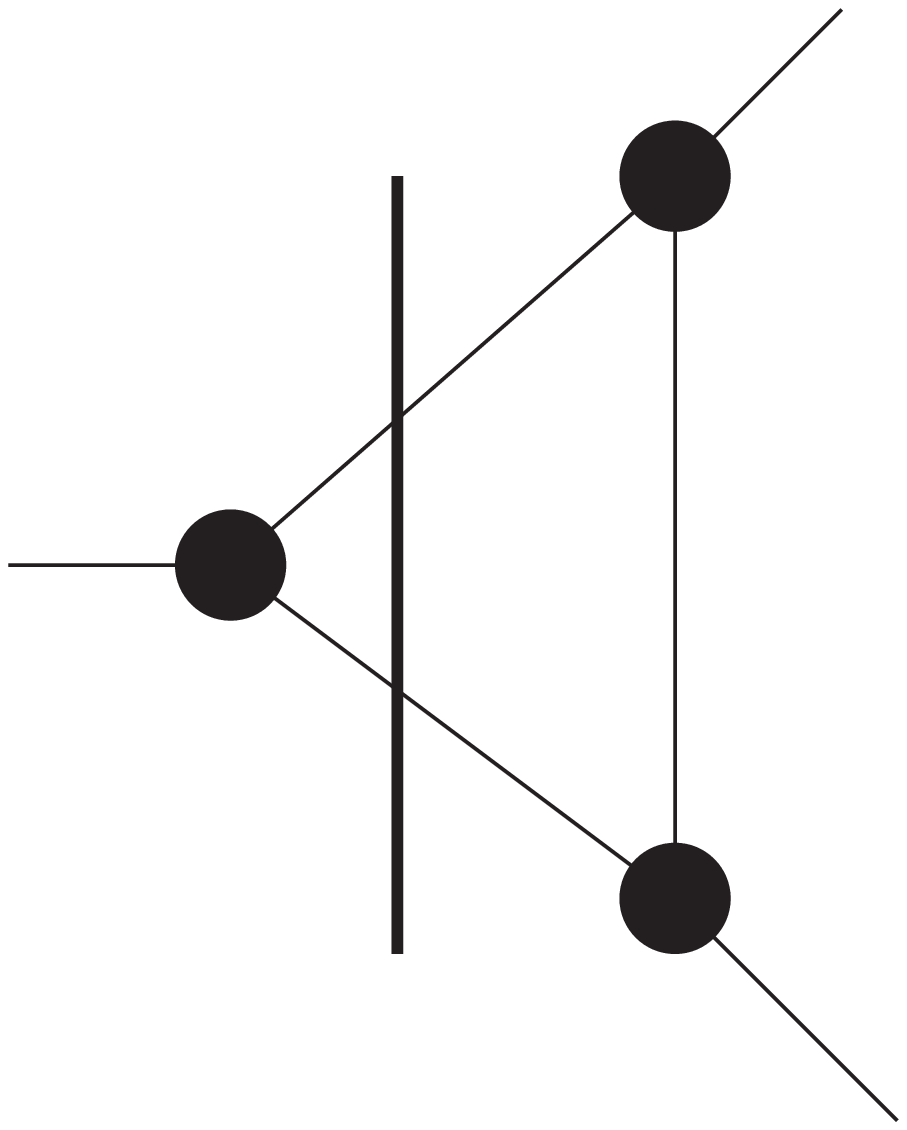}}\;}}
\def\trianglebc{{\;\raisebox{-3mm}{\epsfysize=6mm\epsfbox{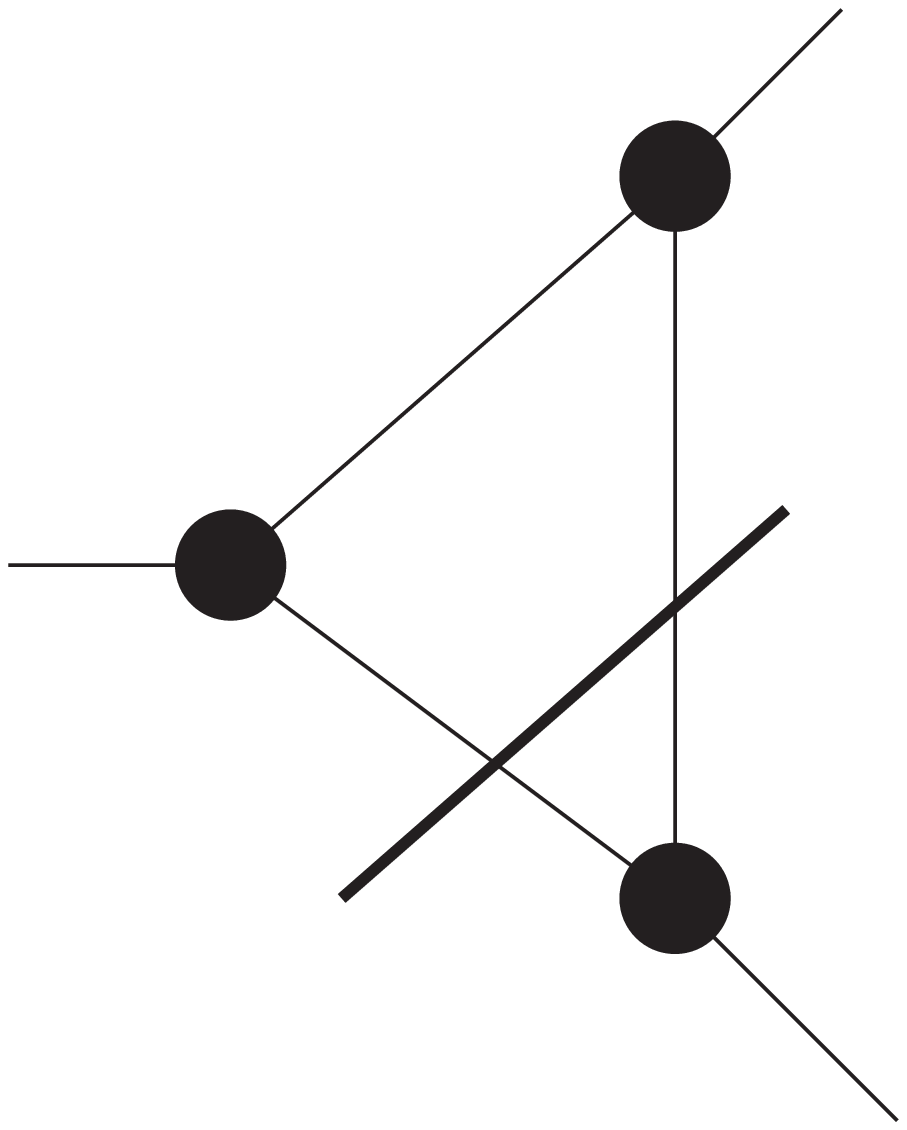}}\;}}
\def\triangleca{{\;\raisebox{-3mm}{\epsfysize=6mm\epsfbox{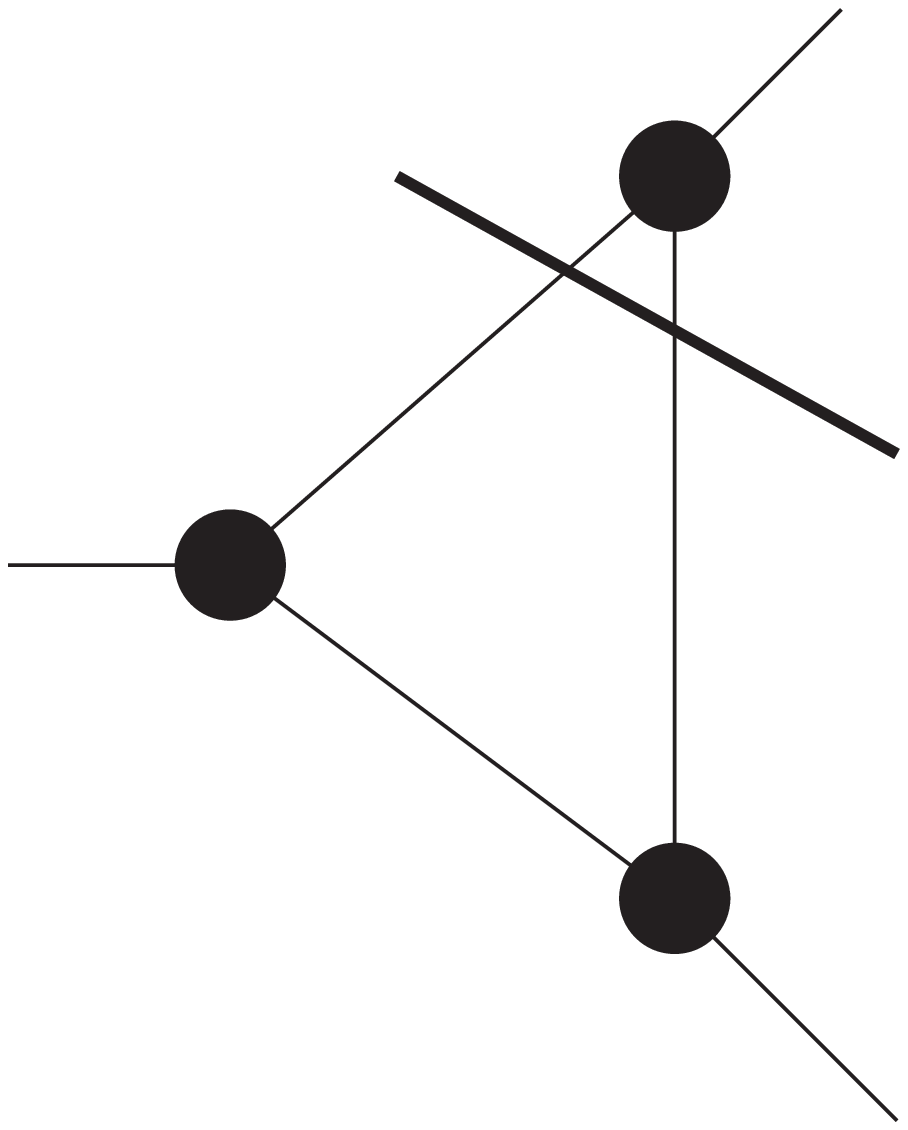}}\;}}
\def\triangleabc{{\;\raisebox{-3mm}{\epsfysize=6mm\epsfbox{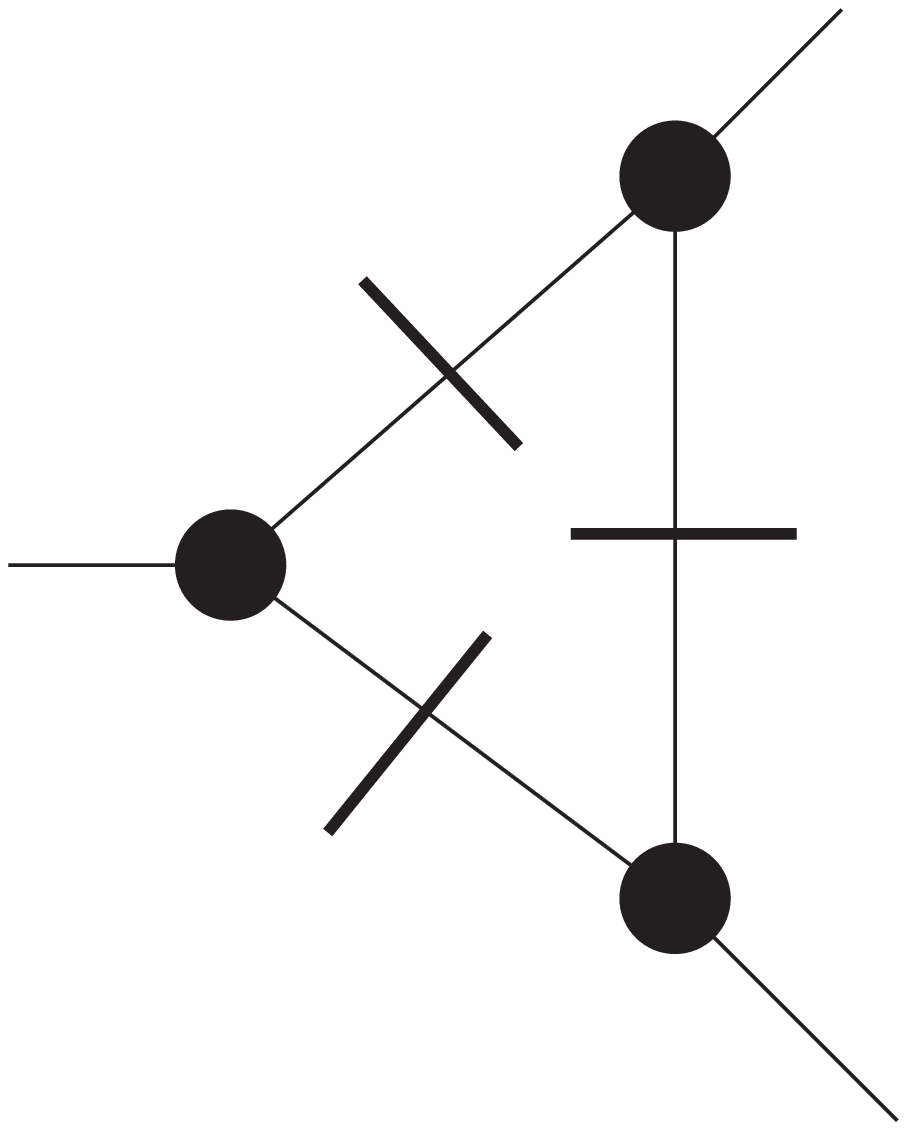}}\;}}

\def\rtra{{\;\raisebox{-3mm}{\epsfysize=6mm\epsfbox{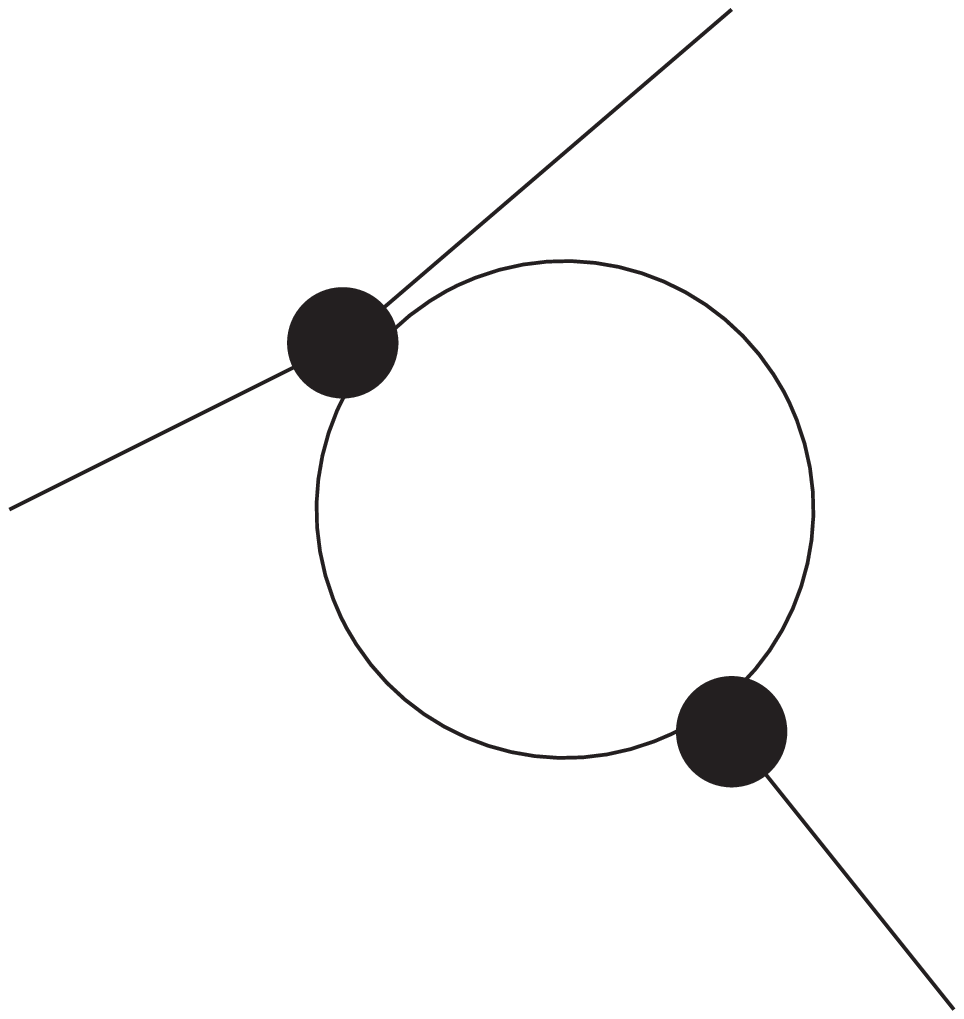}}\;}}
\def\rtrb{{\;\raisebox{-3mm}{\epsfysize=6mm\epsfbox{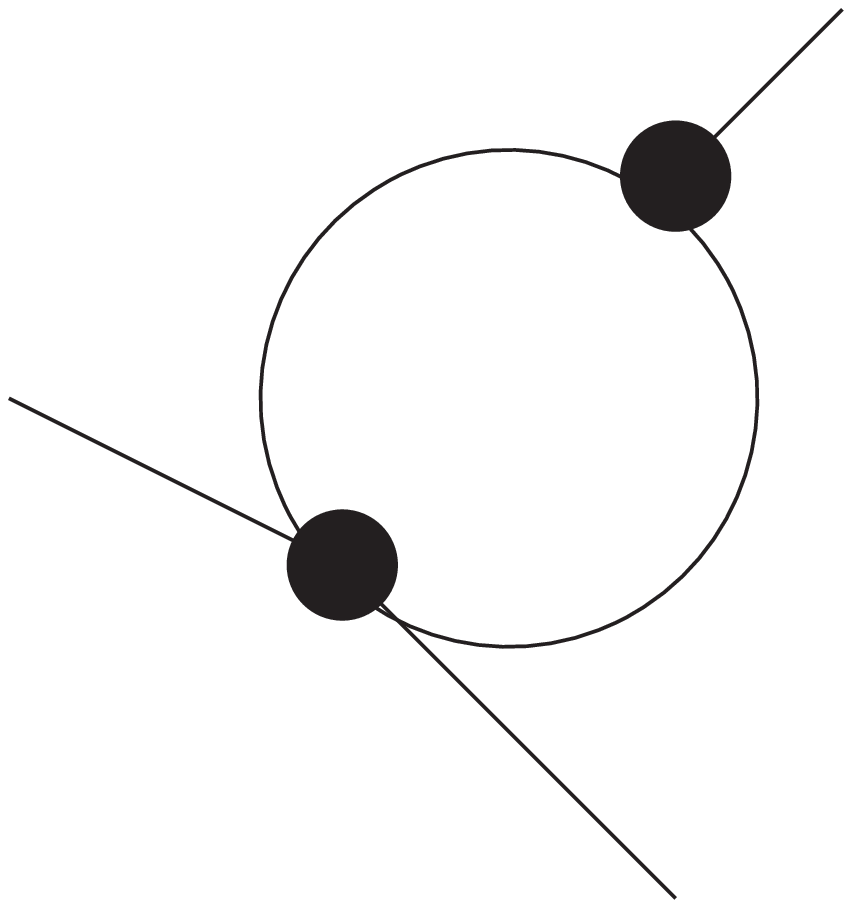}}\;}}
\def\rtrc{{\;\raisebox{-1mm}{\epsfysize=4mm\epsfbox{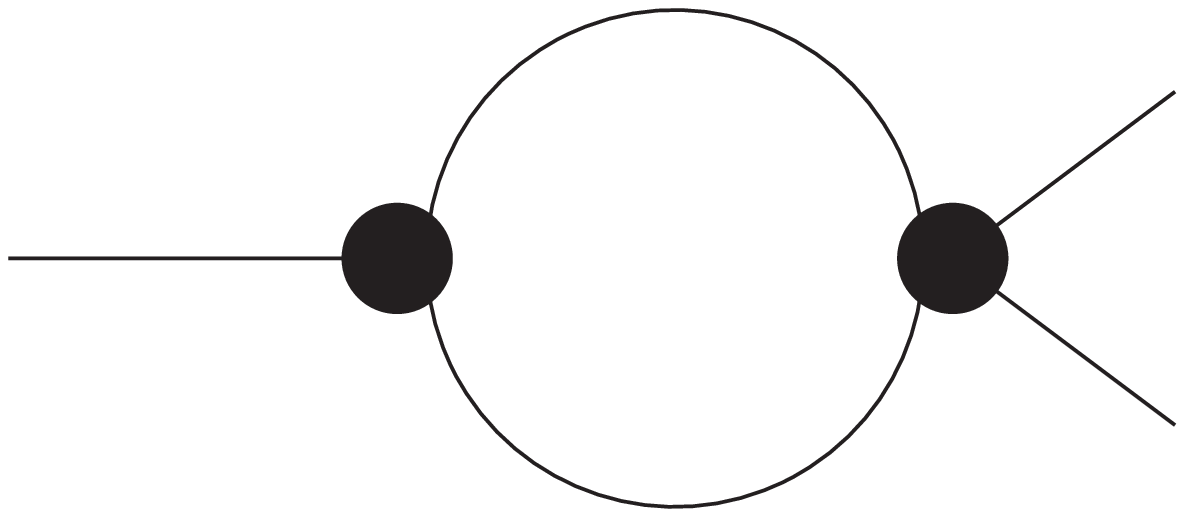}}\;}}

\def\rtrbc{{\;\raisebox{-3mm}{\epsfysize=6mm\epsfbox{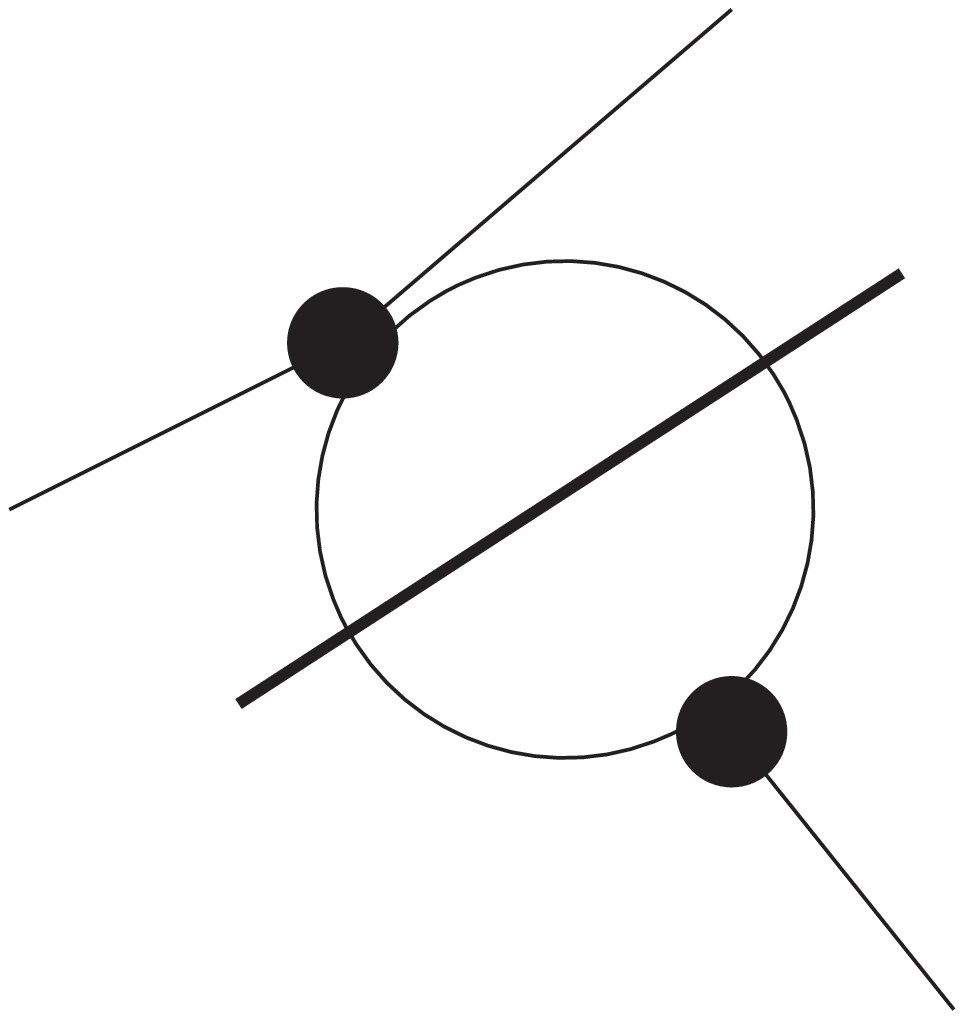}}\;}}
\def\rtrca{{\;\raisebox{-3mm}{\epsfysize=6mm\epsfbox{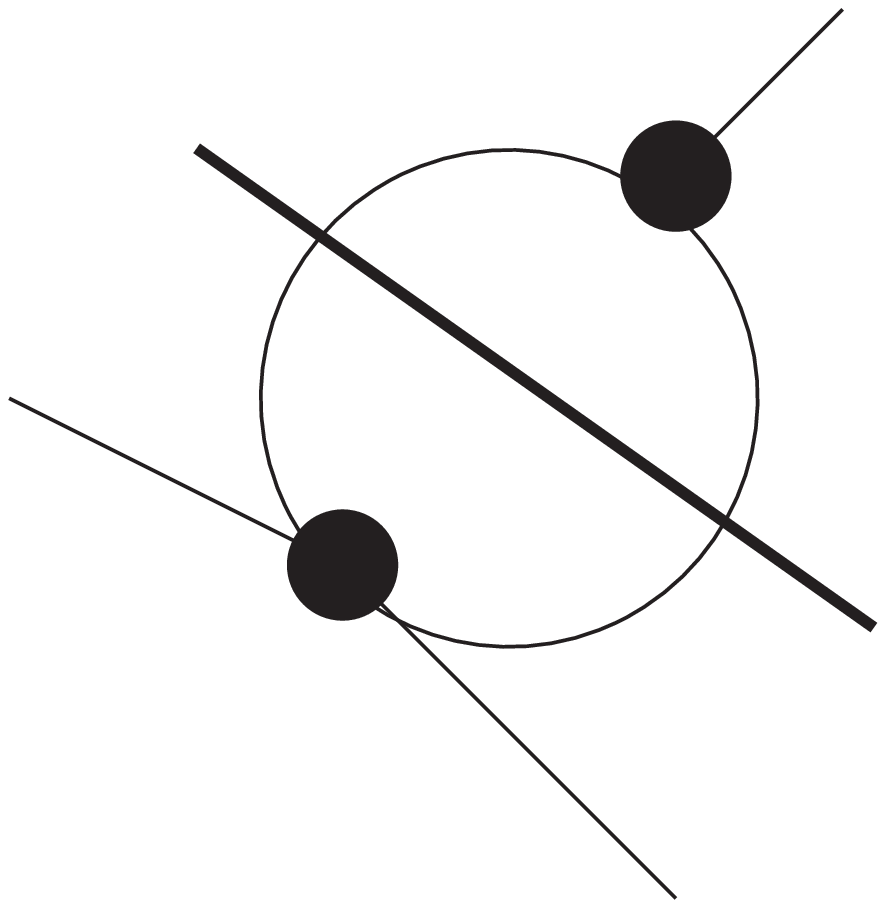}}\;}}
\def\rtrab{{\;\raisebox{-3mm}{\epsfysize=5mm\epsfbox{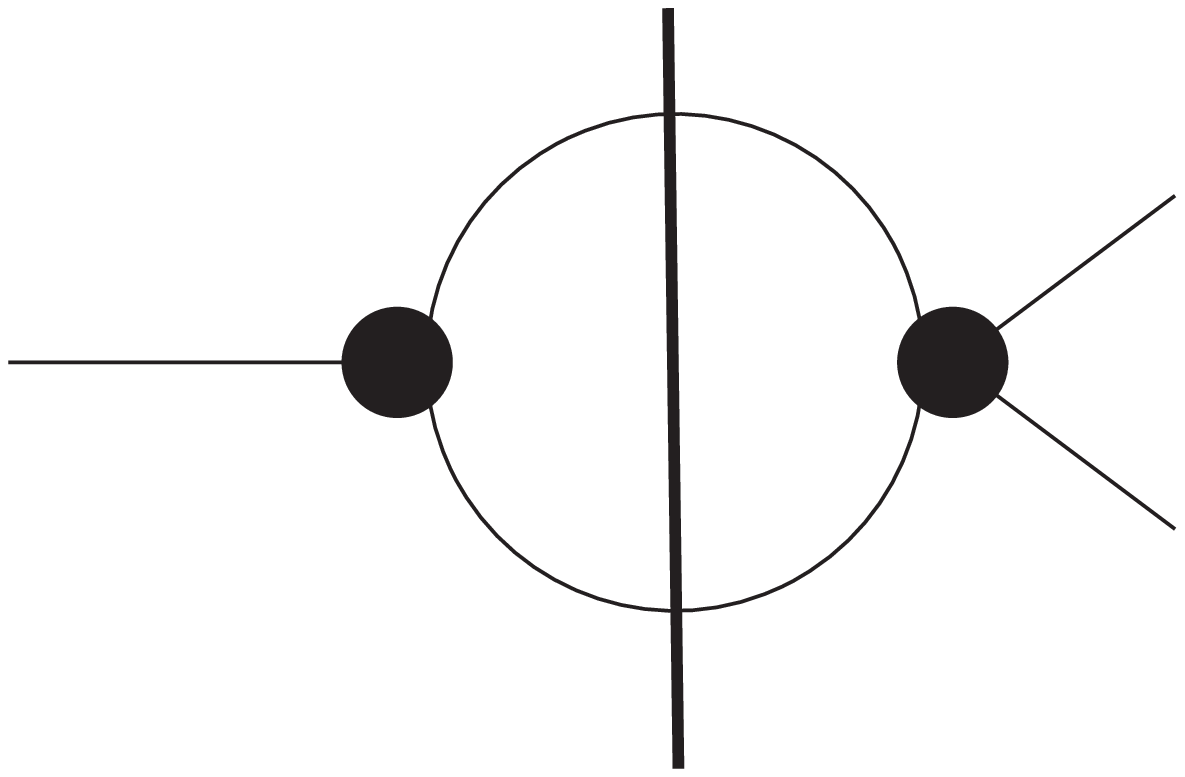}}\;}}

\def\qcd{\;\raisebox{-8mm}{\epsfysize=25mm\epsfbox{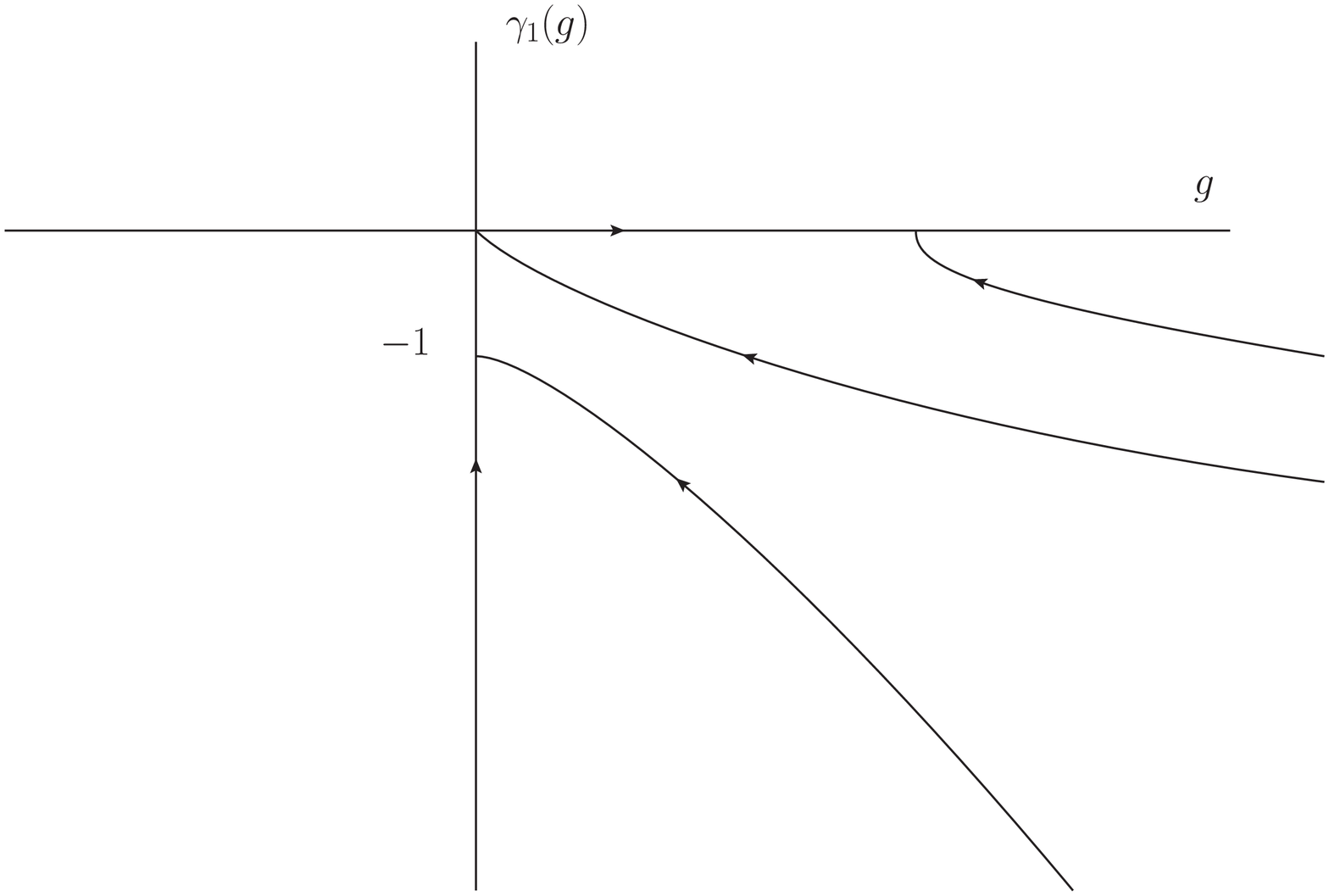}}\;}
\def\qed{\;\raisebox{-8mm}{\epsfysize=32mm\epsfbox{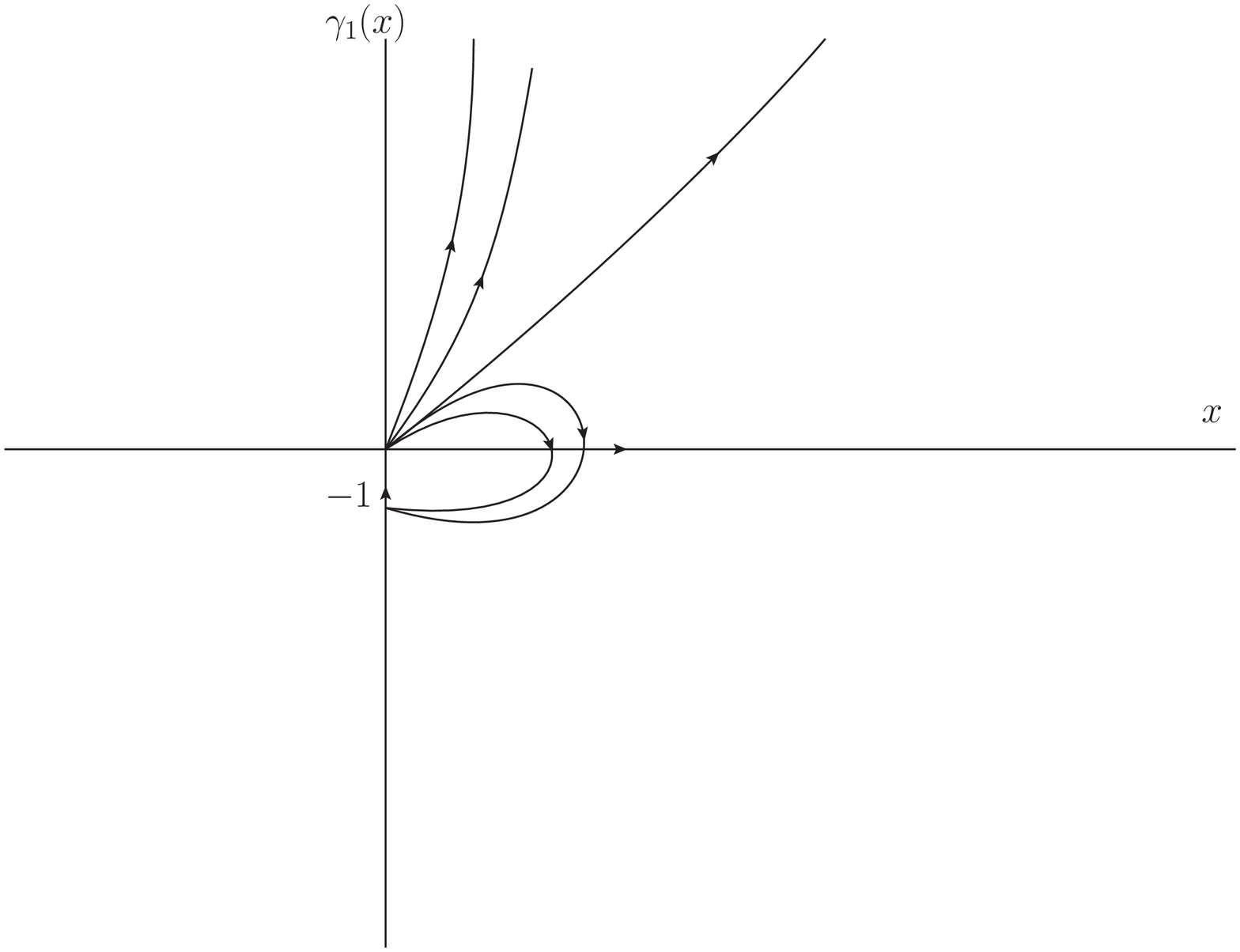}}\;}

\newcommand{\AmS}{{\protect\the\textfont2
  A\kern-.1667em\lower.5ex\hbox{M}\kern-.125emS}}

\title{Algebraic Structures in local QFT}

\author{D. Kreimer\address[IHES]{Institut des Hautes Etudes Scientifiques, 
        35 rte. de Chartres, \\ 
        91440 Bures sur Yvette, France}%
        \thanks{Work supported in parts by grant NSF/DMS-0603781 at Boston University.}}
       
\begin{document}

\begin{abstract}
\vspace{1pc}
\end{abstract}

\maketitle

\section*{Acknowledgments}
{\small It is a pleasure to thank the participants of the Les Houches school on {\em 
Structures in local quantum field theory} \cite{Houches}, Les Houches June 07-25 2010,
where many of the results summarized here were presented in detail, 
for a great and stimulating three weeks. And similarly, thanks to the Loops and Legs community \cite{LL},
for the constant stimulus which nurtures any investigation into the inner workings of quantum fields.}
\section{The polylog as a Hodge structure}
It is the purpose of this short review to emphasize 
the (limiting) mixed Hodge structures underlying quantum field theory \cite{BlKr1,BlKr2,BEK}.
We do so by comparison with the well understood structure of the polylogarithm \cite{Bloch}.

We emphasize that the latter has a representation as an iterated integral, 
and hence also an underlying Hopf algebra structure, see \cite{KrNumbers}.
Indeed, consider the three columns $(C_1,C_2,C3)$ which form a matrix
{\small \be
\left(\begin{array}{ccc}
1 & 0 & 0\\
-\mathrm{Li}_1(z) & 2\pi i & 0\\
{-\mathrm{Li}_2(z)} & 2\pi i \ln(z) & (2\pi i)^2
\end{array}\right).\label{polyH}\ee}
This Hodge structure, apparent in the interplay of colums which compensate their variations amongst each other
(here, a mathematicians variations are a physicists discontinuities along branch cuts)
allows the definition of an invariant
\be \mathrm{Var}(\Im Li_2(z)-\ln|z|\,\Im Li_1(z))=0.\ee
The Hodge sructure and the  Hopf algebra structure cooperate: 
If we define $L(t_k)=\mathrm{L}_k(z)\equiv \ln^k z/k!$, $Li(t_k)=\mathrm{Li}(z)$ 
with  Hopf algebra structure the free commutative algebra on generators 
$t_k$ with coproduct $\Delta(t_k)=\sum_{j=0}^k t_j\otimes t_{k-j}$,
then 
the columnwise branch cut ambiguities compensate in
a combination familiar to a physicist from Bogoliubov's R-operation \cite{KrNumbers}:
$$ \mathrm{Var}\left(m(L^{-1}\otimes Li\circ P)\Delta(t_k)\right)=0.$$
 
Moreover, Griffith transversality leads to a  differential equation which uniquely defines the dilogarithm
from $\mathrm{L}_1(z)$ and $\mathrm{Li}_1(z)$, and 
similar for the higher polylogs.

Let us now return to Feynman graphs.
\section{Hopf algebra of graphs} 
This is a well-studied subject by now \cite{reviews}, so we just list the main formulae.
The Hopf algebra is a free commutative algebra, graded by the loop number:
$H=\mathbb{Q}1\oplus \bigoplus_{j=1}^\infty H^j=\mathbb{Q}1+PH$ 
with a projector $P$ into the augmentation ideal. It hence furnishes:\\
i) the coproduct:
\be\Delta(\Gamma)=\Gamma\otimes 1 + 1 \otimes \Gamma+\overbrace{\sum_{\gamma=\cup_i\gamma_i,\omega_4(\gamma_i)\geq 0}\gamma\otimes\Gamma/\gamma}^{\Delta^\prime(\Gamma)}\ee
ii) the antipode:
\be S(\Gamma)=-\Gamma-\sum S(\gamma)\Gamma/\gamma=-m(S\otimes\mathrm{P})\Delta\ee
iii) the character group:
\be  G^H_V \ni\Phi \Leftrightarrow \Phi:H\to V, \Phi(h_1 h_2)=\Phi(h_1)\Phi(h_2)\ee
iv) the counterterm (parametrized by a Rota--Baxter map $R$):
\bea  S_R^\Phi(\Gamma)& = & -R\left(\Phi(h)-\sum S_R^\Phi(\gamma)\Phi(\Gamma/\gamma) \right)\nonumber\\
& = & -R\;\Phi
\left(m(S_R^\Phi\otimes\Phi\;P)\Delta(\Gamma)\right) \eea 
v) the renormalized Feynman rules:
     \be \Phi_R=m(S^\Phi_R\otimes\Phi)\Delta=[\mathrm{id}-R]m(S_R^\Phi\otimes\phi P)\Delta.\ee
\section{An Example}
The co-product ($\Delta^\prime$ indicates projection into $PH$ on both sides):
\beas \Delta^\prime\left(\verttl\right) & = & 3\vertol\otimes\vertol\\  
+2\feol\otimes\vertol+\phol\otimes\vertol. & & \eeas
The counterterm:
\beas S_R^\Phi(\verttl)  = & & \\ -R m \left[S_R^\Phi\otimes \Phi P \right]\times & & \\
\times \Delta\left(\verttl\right) & & \\
 = -R\left\{\Phi\left(\verttl\right)+\right. & & \\ 
\left.+R\left[\Phi\left(3\vertol+2\feol+\phol\right)\right]\Phi\left(\vertol\right)\right\} & & 
\eeas
The renormalized result:
\beas \Phi_R  =  (\mathrm{id}-R)m(S_R^\Phi\otimes \Phi P)\times & & \\
\times \Delta\left(\verttl\right) & & \\
 =  (\mathrm{id}-R)\left\{\Phi\left(\verttl\right)\right. & & \\
 \left. +R\left[\Phi\left(3\vertol+2\feol+\phol\right)\right]\Phi\left(\vertol\right)\right\}. & & 
\eeas
\section{sub-Hopf algebras}
As a commutative graded Hopf algebra, $H$ has a Hochschild cohomology \cite{ConnKr}. 
This leads to the study of sub-Hopf algebras arising from linear 
combinations of generators contributing at a given order. 
We let $r$ indicate an amplitude from the set $\mathcal{R}$ needing renormalization.
Summing order by order:
\beas c_k^r=\sum_{|\Gamma|=k,\mathrm{res}(\Gamma)=r}\frac{1}{\mathrm|\mathrm{Aut}(\Gamma)|}\Gamma & & \\
\Rightarrow { \Delta(c_k^r)=\sum_j\mathrm{Pol}_j(c_m^s)\otimes c^r_{k-j}.}\eeas
The relevant polynomials $\mathrm{Pol}_j$ can be easily obtained, see \cite{Anatomy,KrvS,WLect} for example.\\
Hochschild closedness says ${bB_+^{r;j}=0}$. Hence,
\be \Delta B_+^{r;j}(X)=B_+^{r;j}(X)\otimes 1 + (id\otimes B_+^{r;j})\Delta(X).\ee
It reduces the sum over all graphs to a sum over primitives (which does not improve the asymptotics that much
after all \cite{DavidAsym})
$$ X^r=1\pm \sum_j c_j^r \alpha^j=1\pm \sum_j \alpha^j B_+^{r;j}(X^r Q^j(\alpha)),$$
\be Q^j=\frac{X^v}{\sqrt{\prod_{\mathrm{edges\; e\; at\; v}}X^e}}.\ee
The latter evaluates to the Lorentz invariant charge
under the renormalized Feynman rules.
This set-up implies locality of counterterms upon application of Feynman rules 
$\Phi B_+^{r;j}(X)=\int d\mu_{r;j}\Phi(X)$:
$$ { \bar{R}(\Gamma)=m(S^R_\Phi\otimes\Phi P))\Delta B_+^{r;j}=\int d\mu_{r;j}\Phi_R(X),}$$
from the co-radical filtration and Hochschild cohomology of $H$.
{
\section{Symmetry}
The study of such sub-Hopf algebras is significant to understand internal symmetry. 
Ward and Slavnov--Taylor identities appear naturally in this context as co-ideals. Combinations 
of graphs as
\be i_k:=c_k^{\bar{\psi}\psi}+c_k^{\bar{\psi}A\!\!\!/\psi}\ee 
span Hopf (co-)ideals $I$:
\be \Delta(I)\subseteq H\otimes I+I\otimes H.\ee
For example, in QED to two loops,
\beas \Delta(i_2) & = & i_2\otimes 1+1\otimes i_2+(c_1^{\frac{1}{4}F^2}+c_1^{\bar{\psi}A\!\!\!/\psi}+i_1)\\
& & \otimes i_1
   +i_1\otimes c_1^{\bar{\psi}A\!\!\!/\psi}. \eeas 
Feynman rules vanish on $I$: $\Phi^R_L(\{\Theta\})(\iota)=0, \forall \iota\in I$
$\Leftrightarrow$ Feynman rules respect quantized symmetry: ${\Phi_R:H/I\to V}$.

Note that this vanishing on $I$ implies the classical equation of motion for the tree-level terms
which we obtain from shrinking internal edges.
Ideals for Slavnov--Taylor identities are generated by equality of renormalized charges \cite{Anatomy,WalterQCD}.
Indeed. the whole set-up for the master equation in Batalin-Vilkovisky goes through
(see Walter van Suijlekom's work \cite{WLect}).
For generalizations of this set-up to Hopf algebras treating all 1PI subgraphs as sub-objects, see \cite{KrvS}.
This has in particular bearing on the the BCFW recursion and quantum gravity, 
which leave much to explore for the future.
}
{
\section{Kinematics and Cohomology}
The above Hochschild cohomology will not only generate the correct perturbation expansion, 
symmetry factors included, but has consequences in understanding scattering, 
upon understanding the role of exact one-cocycles.

Exact co-cycles $b\phi^{r;j}$, one for each closed cocycle $B_+^{r;j}$, are images of the Hochschild coboundary 
operator $b,b\circ b=0$, of maps $\phi^r;j:H\to\mathbb{C}$.
The equivalence class is hence described as 
\be [B_+^{r,j}]=B_+^{r;j}+b\phi^{r;j}=\tilde{B_+}^{r;j}.\ee
It has its natural role in describing the variation of scattering angles in renormalized amplitudes,
using dedicated such  $\phi^{r;j}: H\to \mathbb{C}$ for each generator of the cohomology class.
Indeed, we look at the variation of external momenta separated as a variation of an overall scale and 
angles defined by that scale:
\be G_R^r(\{g\},\ln s,\{\Theta\})=1\pm \Phi_R^{\ln s,\{\Theta\}}(X^r(\{g\}))\ee
with $X^r=1\pm \sum_j g^j B_+^{r;j}(X^r Q^j(g))$, $bB_+^{r;j}=0$,
and  \be G^r_R=\left[\sum_{j=1}^\infty \gamma_j^r(\{g\},\{\Theta\})\ln^j s\right]
+\overbrace{G_0^r}^{{abelian\; factor}}.\ee
Here, the abelian factor summarizes all terms which contribute to overall convergent formfactors
in the Green function. They all can be subsumed by extending the group $\mathrm{Spec}(H)$ by an abelian factor,
corresponding to the cocommutative and commutative Hopf algebra of convergent graphs.

Then, for MOM and similar schemes (not MS!):\\
 \be  \{\Theta\}\to\{\Theta^\prime\} \Leftrightarrow B_+^{r;j}\to B_+^{r,j}+b\phi^{r,j},
\ee 
describes the variation of angles at a fixed scale $L$ through the addition of an exact term, while 
 \be  \Phi^R_{L_1+L_2,\{\Theta\}}=\Phi^R_{L_1,\{\Theta\}} \star \Phi^R_{L_2,\{\Theta\}}.\ee 
generates 1-parameter groups of automorphism coming from rescaling at fixed angles.
} 
{
\section{Leading log expansions and the RGE}
For each vertex $v$, we have a combinatorial charge $Q^v$:
\be Q^v(g)=\frac{X^v(g)}{\prod_{e}\sqrt{X^e}},\ee
$e$ adjacent to $v$. Under the renormalized Feynman rules $\Phi_R$, $Q^v$ evaluates
to a Poincar\'e invariant charge of the theory (it is not invariant under conformal 
transformations - charges run when dilated). Using the underlying Hopf algebra structure, the renormalization group equation 
(RGE)
\be \left( \partial_L +\beta(g)\partial_g-\sum_{e\;adj\;r}\gamma_1^e\right) G^r(g,L)=0\ee
rewrites in terms of the Dynkin operator {(}$\gamma_1^r(g)=S\star Y (X^r(g))${)}:
\be \gamma_k^r(g)=\frac{1}{k}\left(\gamma_1^r(g)-\sum_{j\in R}s_j\gamma_1^jg\partial_g\right)\gamma_{k-1}^r(g),\label{grek}\ee
as the expected recursion for the leading log expansion \cite{KrYeats6,YThesis}.
}
{
\section{Ordinary differential equations vs DSE}
The series over all graphs $X^r$ contributing to an amplitude $r$ was given above as a solution to a fixpoint
equation in Hochschild cohomology. Under $\Phi_R$, this fixpoint equation turns into the familiar 
Dyson--Schwinger equations (DSE), and the distinguished primitive 
elements $B_+^{r;j}(1)$ turn into the skeleton integral kernels
for these equations. It is a  remarkable fact that the resulting 
sub-Hopf algebras are invariant on any finite truncation 
of the sum over skeletons \cite{BergbKr}.
Using the iterated integral  structure (see \cite{BrownIt}
for a formal exposition of iterated integrals) 
\be \Phi_R(B_+^{r;j}(X))=\int \Phi_R(X)d\mu_{r;j}\ee
allows to combine $X^r=1\pm \sum_j B_+(X^rQ^j)$ with the RGE to a system of ordinary differential equations
$\forall r\in \mathcal{R}$
\be \gamma_1^r=P(g)-[\gamma_1^r(g)]^2+\sum_{j\in \mathcal{R}}s_j\gamma_1^j g\partial_g\gamma_1^r(g),\ee
for the linear term in the leading log expansion, whilst the higher terms are determined as before,
see (\ref{grek}).

We have a single equation for the $\beta$-function in massless gauge theories:
$\beta(g)=g\gamma_1(g)/2$, for $\gamma_1$ the anomalous dimension of the massless gauge propagator
\be {\gamma_1(g)}={P(g)-\gamma_1(g)(1-g\partial_g)\gamma_1(g)}.\ee
This uses the Ward identity in  QED and a background field gauge for QCD, see below.

A few words concerning the function $P$, whose existence we assume, are appropriate.
It contains certainly the periods $\mathrm{res}(p)$ which are generated by primitive 1PI graphs $p\in H$
as coefficient of the term linear in $L$, $\Phi_R(p)=\mathrm{res}(p)L$.
What else? To understand this, consider 
two primitives $p_1,p_2$, and assume there are $n_1$ insertion places for $p_2$ in $p_1$, and $n_2$ for $p_1$ in $p_2$.

Define $n_+=(n_1+n_2)/2$, $n_-=(n_1-n_2)/2$. Then, using the pre-Lie insertion of graphs $\star$, 
the symmetric \be n_+\left(\frac{1}{n_1}p_1\star p_2+\frac{1}{n_2}p_2\star p_2-  p_1 p_2\right),\ee
is a primitive element in the Hopf algebra generated by a suitable Dynkin operator, whilst the antisymmetric
\be n_-\left(\frac{1}{n_1}p_1\star p_2-\frac{1}{n_2}p_2\star p_2\right)\ee
also evaluates to only a term linear in $L$, and can best be interpreted as the dual of the normalized commutator
\be [Z_{p_1},Z_{p_2}]\in L\ee in the Lie algebra $L$ dual to $H$.
The primitives $p$, and their symmetric extensions above, correspond all to elements in the center
of the Lie algebra $L$ appearing in $H=U^\star(L)$ by Milnor Moore. 
We thus expect that a dedicated study of the lower central series of 
said Lie algebra is mandatory to understand $P$ fully, 
to incorporate the commutator subgroups correctly.
}

\section{Limiting mixed Hodge structures}
The Hopf algebra $H$ can be obtained from flags
\be f:=\gamma_1\subset\gamma_2\subset\ldots\subset\Gamma,\;\Delta^\prime(\gamma_{i+1}/\gamma_i)=0.\ee
The set of all such flags $F_\Gamma\ni f$ determines the  Hopf algebra structure.
We let $|F_\Gamma|$ be the length of the flag, 
while the number of flags is determined from the decomposition into maximal forests, 
and hence is sensitive to the number of overlapping subdivergences \cite{BlKr1}.

The flag decomposition then determines a column vector $v=v(F_\Gamma)$ and a nilpotent 
matrix ${(}N{)}={(}N({|F_\Gamma|}){)}$, 
${(}N{)}^{k+1}=0$, $k=\mathrm{corad}(\Gamma)$ such that 
 \beas \lim_{t\to 0}{(}e^{- \ln t {(}N{)}}{)}
\Phi_R(v(F_\Gamma))= & & \\
(c_1^\Gamma(\Theta)\ln s,c_2^\Gamma(\Theta),c_k^\Gamma(\Theta)\ln^k s)^T & & \eeas
where $k$ is determined from the co-radical filtration and $t$ 
is a regulator say for the lower boundary in the parametric representation.

The vector on the rhs is a vector of periods in the mathematical sense.
Renormalization hence appears as a limiting mixed Hodge structure.
In fact, \cite{BlKr1} contains a full treatment for arbitrary internal masses and external momenta of
that limit, leading to a renormalized integrand which is a well-defined projective differential form
with a log-homogenous integrand. This opens renormalized amplitudes to the same study by algebro-geometric methods, as
recently applied to the periods provided by primitive graphs.

The study of these periods actually initiated the discovery of the Hopf algebra structures underlying QFT here reported. It started 
in collaboration with David Broadhurst \cite{BroadKr9}, and after the connection to algebraic geometry
was established in \cite{BEK}, the study of these periods recently blossomed thanks to the works of 
Francis Brown \cite{Brown1,Brown2}, Dzmitry Doryn \cite{Doryn1,Doryn2}, 
Oliver Schnetz \cite{SchnC,BrSchn} and Karen Yeats \cite{BrY}.
The progress is mainly due to a detailed study of the linear algebra and algebraic geometry of the two Kirchhoff polynomials,
which also leads to an improved understanding of Green's function as functions of 
internal masses and external momenta \cite{BlKr2}.
{
\section{The Feynman graph as a Hodge structure} 
Let us dwell a moment on the last point.
In general, to learn about the analytic structure of Feynman integrals as functions of external momenta and internal masses,
we start from a Hopf algebra structure as above. Here we exhibit the simplest case of 
one-loop graphs, which provide the simplest primitives of $H$. Actually, we focus on the one-loop triangle:
{\small $$
\left(\begin{array}{ccccc}
1 & 0 & 0 & 0 & 0\\
\rtrc & \rtrab  & 0 & 0 & 0\\
\rtrb & 0 & \rtrca  & 0 & 0\\
\rtra & 0 & 0 & \rtrbc  & 0\\
{\triangle} & \triangleab & \triangleca & \trianglebc & \triangleabc
\end{array}\right),
$$}
which provides five column vectors $(C_1,C_2,C_3,C_4,C_5)$, say.
In the first column, the first entry $1$ corresponds to the tree level term.
Then, the next three entries are the three reduced diagrams which we can assign to the one-loop triangle graph.
The fifth entry is finally the triangle itself. Columns 2,3,4 then put suitable internal propagators on the mass shell, hence
capturing suitable variations of the entries in the first column. 

Note that the two cuts in the reduced diagrams 
leave no free integration, but give a mass and momenta dependent constant which generalizes the entry $2\pi i$ in the 
Hodge structure of the dilog in Eq.(\ref{polyH}). The triangle cut at two internal edges leaves one free integration of the form 
$\int dz/(az+b)$, corresponding to the entry $\ln z$ before. 

Finally, the last column has three internal propagators on the mass shell. It hence exceeds what Cutkosky would tell us,
and closes the above matrix to a Hodge matrix by the results of \cite{BlKr2}. 
The recent physicists practice to put more internal propagators on the mass shell than prescribed by Cutkosky has indeed, it seems,
a nice  mathematical origin, see \cite{BlKr2}. 
It is tempting to muse that many questions regarding cut-reconstructability could be clarified 
by clarifying the Hodge structure underlying quantum field theory.

So for the future: what, then, can we learn from {\em invariant triangles} as in 
{$$\mathrm{Var}\left(\Im \triangle-\left[\Re \trianglebc \cdot \Im \rtra
\right] +\cdots\right)=0,$$}
considered as functions of complex external momenta and internal masses? 
}
{
\section{QED}
We now turn to aspects of resummation of Feynman diagrams, and hence to  rather modest 
first attempts to connect the Hodge and Hopf algebra approach exhibited here to non-perturbative physics. 
Lacking precise knowledge of the generalized skeleton function $P(x)$, we study what can be said
as a function of assumptions on $P(x)$. We start with QED \cite{vBKUY1}. 
It suffices to consider the sub Hopf algebra for vacuum polarization graphs.
We are led to a single ODE:
\be \gamma_1(x)=P(x)-\gamma_1(x)^2+\gamma_1(x)x\partial_x\gamma_1(x),\ee   with $P(x)>0$ and 
$P(x)$ twice differentiable assumed. Also, at some small coupling $x_0$, we assume 
$\gamma_1(x_0)=\gamma_0>0$.\\
We then find different solutions distinguished by flat $e^{-\frac{1}{x}}$ behaviour but with 
an identical perturbation expansion near the origin. Switching to the coupled system with the running charge included,
$\frac{d\gamma_1}{dx}=\gamma_1-\gamma_1^2-P$, $\frac{dx}{dL}=x\gamma_1$,
$L=\int_{x_0}^{x(L)}\frac{dz}{z\gamma_1(z)}$,
we find three types of solutions with a separatrix separating those above which all have a Landau pole, 
from those below which are double valued and hence unphysical, see Fig.(\ref{qed}).
\begin{figure}
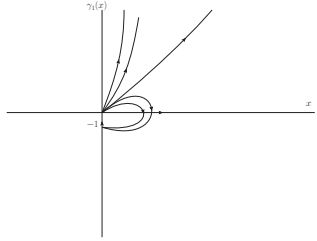
$$\qed$$\caption{QED: The $\beta$-function ($\beta=\gamma_1/2$) as a function of the coupling.
The separatrix is the only solution which might have no Landau pole.}\label{qed}\end{figure}
The separatrix exists under rather mild conditions on $P(x)$
and might or might have not a  Landau pole, as a function of much finer asymptotics of $P(x)$
for large coupling $x$. See Karen Yeats talks at \cite{Houches}, or see \cite{vBKUY1,vBKUY2}, for details.
}
{
\section{QCD}
Again, the sub Hopf algebra for gluon polarization graphs suffices, now upon using the 
 background field gauge \cite{vBKUY2}. We have the same ODE,
\be \gamma_1(g)=P(g)-\gamma_1(g)^2+\gamma_1(g)g\partial_g\gamma_1(g),\ee   with $P(g)<0$.
We assume $P(g)$ twice differentiable and concave near 0. 
Now, the separatrix is indeed clearly the solution preferred by physics: 
its the unique solution which flows into $(0,0)$ at large $Q^2$.
The scale of QCD is then set by 
$L=\int_{g_0}^{g(L)}\frac{dz}{z\gamma_1(z)} \to L_\Lambda=
-\int_{g(L_\Lambda)}^\infty\frac{dz}{z\gamma_1(z)}$,\\ $L_\Lambda=\ln Q^2/\Lambda_{QCD}$.\\
For $P(g)<0$, the picture is given in Fig.(\ref{qcd}).
\begin{figure}
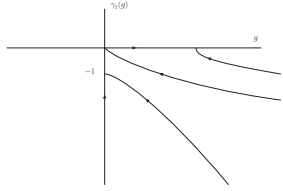
$$\qcd$$\caption{QCD: The $\beta$-function ($\beta=\gamma_1/2$) as a function of the coupling. Only the separatrix flows into the origin
at high energy.}\label{qcd}\end{figure}
The separatrix exists and gives an asymptotic free solution, with a finite mass gap possibly for the inverse propagator
iff $\gamma_1(x)<-1$ for some $x>0$.  See \cite{vBKUY2} for technical details using
for dispersive methods as introduced by Shirkov et.al.\ in field theory.
}

\end{document}